\documentclass[prb,reprint]{revtex4-1}

\usepackage{amsmath}  
\usepackage{amsfonts} 
\usepackage{graphicx} 

\newcommand{\be}{\begin{equation}}
\newcommand{\ee}{\end{equation}}
\newcommand{\bse}{\begin{subequations}}
\newcommand{\ese}{\end{subequations}}
\newcommand{\bea}{\begin{eqnarray}}
\newcommand{\eea}{\end{eqnarray}}

\newcommand{\comment}[1]{}

\begin{document}


\title{Space symmetries draw elasticity theory}

\author{Chaouqi Misbah}
\email{chaouqi.misbah@ujf-grenoble.fr}
\affiliation{Universit\'e Grenoble I/CNRS, Laboratoire Interdisciplinaire de Physique/UMR5588, Grenoble F-38041, France}
\author{Sofia Biagi}
\email{sofia.biagi@ujf-grenoble.fr}
\affiliation{Universit\'e Grenoble I/CNRS, Laboratoire Interdisciplinaire de Physique/UMR5588, Grenoble F-38041, France}
\author{Paolo Politi}
\email{paolo.politi@cnr.it}
\affiliation{Istituto dei Sistemi Complessi, Consiglio Nazionale delle Ricerche, Via Madonna del Piano 10, 50019 Sesto Fiorentino, Italy}


\date{\today}

\begin{abstract}
The foundation  of continuum elasticity theory  is based on two general principles: (i) the force felt by a small volume element  from its surrounding acts only through  its surface (the Cauchy principle, justified by the fact that interactions are of short range and are therefore localized at the boundary);
(ii) the stress tensor must be symmetric in order to prevent spontaneous rotation of the material points. These two requirements are presented to be necessary in classical textbooks on elasticity theory.
By using only basic spatial symmetries it is shown that elastodynamics equations can be derived,
for high symmetry crystals (the typical case considered in most textbooks),
without evoking any of the two above physical principles.
\end{abstract}

\maketitle 

\comment{
\pacs{{87.16.D-} 
{83.50.Ha} 
{87.17.Jj} 

{83.80.Lz} 
{87.19.rh} 
}
}

\section{Introduction}
When teaching elasticity theory one is faced with the question of whether it is necessary to introduce series of postulates (to be recalled below) in order to arrive to the final elastodynamics equations (like Hookes law, Lam\'e equation, etc...), or is there a simpler way bypassing general principles (except using the Newton law). Moreover, traditionally, very quickly the notion of tensors (of second and fourth rank, see below) is used, which adds new obstacles to many students. Here we give a direct derivation of elasticity equations by evoking only elementary space symmetries. The notion of stress tensor (of second rank) appears only at the very end as a natural consequence of elementary symmetries, and not as an initial notion entering the ingredients of the theory. The main spirit of this paper  is presented in section II.

Let us first recall the traditional foundation principles of elasticity theory.
In classical textbooks, such as those of Landau and Lifchitz \cite{Landau},
Feynman \cite{Feynman}, and others~\cite{Love,Timoshenko,Kittel},
this theory is based on the following three steps.

(i) {\it The force felt by a given material element acts only on the surface.}
Consider a small volume element of the material having a volume $\Delta V$, enclosed by a surface $\Delta A$. After deformation   that  element would feel a force distribution from the surrounding medium acting only through its surface.  The elementary force $d\mathbf f$ acting on a portion $dA$ of the closed surface $\Delta A$ has the form
\begin{equation}
d\mathbf f= \mbox{\boldmath  $\sigma$} \mathbf n dA ,
\label{eq_Landau}
\end{equation}
where {\boldmath $\sigma$} is the stress tensor and $\mathbf n$ is the  normal unit vector.
That the force is only localized at the surface (and not to the bulk element) is often
justified by the fact that the interactions are of short range.
For example, it is stated in the book of Landau and Lifchitz \cite{Landau} (p.~4):
{\it We can say that the forces which cause the internal stresses are, as regards the theory of
elasticity, ``near-action" forces [...] Hence it follows that the forces exerted on
any part of the body by surrounding parts act only on the surface of
this part.}

The fact that the force must be localized at the surface and have the form (\ref{eq_Landau})
is known as the Cauchy principle. The resultant force applied on the volume element is
obtained upon integration over the surface,
$\mathbf f= \int _{\Delta A} \mbox{\boldmath  $\sigma$} \mathbf n dA
= \int _{\Delta V} \nabla\cdot\mbox{\boldmath  $\sigma$} dV $,
where we have used the divergence theorem. This means that the force per unit volume is simply
given by  $\nabla\cdot\mbox{\boldmath  $\sigma$}$. From this and the Newton law we can write
     \begin{equation}
     \label{newton}
     \rho \ddot {\mathbf u}= \nabla\cdot\mbox{\boldmath  $\sigma$},
     \end{equation}
      where $\rho$ is the mass density, $\mathbf u$ is the displacement vector, and
$\rho \ddot {\mathbf u}$ represents the inertial force per unit volume
(double dots designate  the second time derivative).

In the book of Feynman~\cite{Feynman} the fact that the stress tensor
acts on the surface is justified as follows
(p.~31-6  of Ref.~\onlinecite{Feynman}): {\it Consider a body of some elastic material --say a block of jello. If we make a cut through a block, the material on each side of the cut will, in general, get displaced by internal forces. Before the cut was made , there must have been forces between the two parts of the block that kept the material in place; we can define the stresses in term of these forces.}

(ii) {\it The stress tensor is symmetric.} The stress tensor $\mathbf \sigma$ must be symmetric, i.e.
    \begin{equation}
    \label{sigma}
    \mathbf \sigma_{ij}=\mathbf \sigma_{ji}.
     \end{equation}
     The justification of this condition varies from one book to the other.
In Ref.~\onlinecite{Landau}, calculating the torque of the elastic forces on the volume element, it is stated that if the stress tensor is not symmetric then the total torque would have not only a surface contribution, but also a bulk contribution, which would contradict the fact that the forces are localized to the surface only. The bulk contribution vanishes if the stress tensor is assumed to be symmetric. A variant of this proof can be found in the book of Feynman \cite{Feynman}  where it is said (p.~31-11):
{\it We can also show that $S_{ij}$ is a symmetric [stress] tensor by looking at the
forces on a little cube of material. [...] Now there must be no torque on the cube,
or it would start spinning.}
\comment{All these considerations have physical motivations. It will be shown below that
actually these considerations are the results of space symmetries.}
Similar arguments are given in the books of Love~\cite{Love}, Timoshenko and Goodier~\cite{Timoshenko},
and Kittel~\cite{Kittel}.


(iii) {\it The stress tensor has a linear relationship with the strain tensor.}
The stress tensor has a proportionality relation with the strain tensor
$\epsilon_{ij}={1\over 2}(\partial _ i {\mathbf u}_j+
\partial _j {\mathbf u}_i)$,
where ${\mathbf u}_i$ is the $i$-th cartesian component~\cite{note_notation}
 of the displacement
vector $\mathbf u$.
For an isotropic material, we have
\begin{equation}
\sigma _{ij}= \lambda \epsilon _{kk}\delta_{ij}+ 2\mu \epsilon_{ij},
\label{hooke}
\end{equation}
where $\lambda$ and $\mu$ are the two Lam\'e coefficients.\cite{note_Lame}
In most books (see for example Ref.~\onlinecite{Feynman}, p.~31-7) this is derived from the energy by saying that the elastic energy is a quadratic function of the strain ($F= \Lambda_{ijkl}\epsilon_{ij}\epsilon_{kl}$), by (i) and (ii) evoking the symmetry of the stress tensor, and by using some general mathematical relations obeyed by the rank four tensor $\Lambda_{ijkl}$ for an isotropic medium. Remark in passing, that  our own experience is that many students are quite reluctant  already regarding  stress and strain tensors (second-rank tensors), and this becomes even worse when a fourth-rank tensor is evoked.

\comment{
The above relation (\ref{hooke}) is derived in different ways depending on text books. Both in the book of Landau and Lifchitz and Feynman it is stated that the elastic  energy $E$ depends quadratically on the deformation tensor $\epsilon_{ij}$, and since the energy is a scalar, the most general form of $E$ is $E=C_{ijkl} \epsilon_{ij} \epsilon_{kl}/2$ where $C_{ijkl}$ is a fourth order tensor, having in principle 81 components in there dimensions. Then using several manipulations about symmetry of the crystal, the number of independent coefficients is equal to two. Then use is made of a relation   between energy and the stress tensor, $\sigma_{ij}=\partial E/\partial \epsilon_{ij}$ (the derivation of which requires several steps; see \cite{Landau}), leading to $\sigma_{ij}=C_{ijkl} \epsilon _{kl}$. At this stage $\sigma_{ij}$ is not automatically symmetric, because a priori we have no information on $C_{ijkl}$. !
 Then use is made the fact that the total force applied on the material is a divergence of the stress tensor
}

Reporting (\ref{hooke}) into (\ref{newton})  one arrives to the elastodynamics equations
for an isotropic material,
\begin{equation}
\rho \ddot {\mathbf u}= \mu\nabla ^2 {\mathbf u}+ (\lambda+\mu) \nabla (\nabla\cdot{\mathbf u}) .
\end{equation}

\comment{where E is the Young modulus and $\nu$ is the Poisson ratio and are related to the Lamé coefficients by $E=\mu (3\lambda + 2\mu)/(\lambda + \mu )$ and $\nu =\lambda /2(\lambda + \mu )$.
}


The objective of this article is to show that for high symmetry crystals
(such as an isotropic and cubic crystals in 3D and isotropic, square and hexagonal in 2D crystals) none of the above three points is necessary in order to arrive
to the elastodynamics equations. Rather, evoking only elementary space symmetry is sufficient.

\section{Derivation of elasticity equations from elementary space symmetries}
In order to make the derivation as pedagogical as possible, we shall consider first a two dimensional system. Let $\mathbf u (x,y)$ be the displacement vector having $(u,v)$ as Cartesian components along the $x$ and $y$ directions. An elastic force takes place only if $\mathbf u (x,y)$ is non constant, otherwise all the material element would displace by the same quantity, corresponding to a rigid translation. The fact that $\mathbf u (x,y)$ is non constant is expressed by non zero first derivatives, and possibly higher order derivatives. For the sake of abbreviation, we shall denote the first derivatives as $u_x\equiv \partial u/\partial x$, $u_y\equiv \partial u/\partial y$, and so on. Second derivatives will be denoted as $u_{xy}\equiv \partial ^2u/\partial x \partial y$, $v_{xx} \equiv \partial ^2v/\partial x ^2$, and so on. Thus the forces are functions of derivatives of $u$ and $v$.

Like in most physical theories (elecromagnetism, diffusion,
hydrodynamics,~\cite{note_long_range}...) we shall seek for the lowest derivatives that provide relevant
results. It will turn out that we need to expand the force  up to second derivatives of $\mathbf u (x,y)$. The force (with Cartesian components denoted as $U$ and $V$) acting on a material element is a function of derivatives of $\mathbf u (x,y)$, $U=U( u_{i},v_i, u_{ij}, v_{ij})$ and $V=V( u_{i},v_i, u_{ij}, v_{ij})$.
Higher order derivatives could also be considered if one is interested in the so-called ``second gradient" theory, a step beyond the classical elasticity theory, a question which  is beyond our objective. In two dimensions there are four distinct first derivatives $u_x,u_y,v_x,v_y$ and six distinct second derivatives $u_{xx},u_{xy},u_{yy}, v_{xx}, v_{yy},v_{xy}$. From Newton law, the acceleration is proportional to the force (we consider a force per unit volume for the sake of comparison with traditional theory), and the two components of displacement field obey
\begin{equation}
\mathbf \rho \ddot u= U,\;\;\;
\mathbf \rho \ddot v= V .
\end{equation}

Since we are interested in linear elasticity, the force components $U$  and $V$ are developed in a
Taylor expansion in terms of their arguments by keeping only  linear terms in $u$ and $v$.
Considering the terms with first derivatives, we have explicitly
\be
 \label{equv}
\begin{split}
\mathbf \rho \ddot u &= a u_x+ b u_y+ cv_x + dv_y\\
\mathbf \rho \ddot v &= a' u_x+ b' u_y+ c'v_x + d'v_y
\end{split}
\ee
where the coefficients $a,b,...$ are real constant.
Let us start with the case of an isotropic material. The evolution equations must be separately invariant under each of the following symmetries:
\be
\label{sym}
\begin{split}
& x\rightarrow -x \;  \;\;\; \mbox{and} \;\;\; \;u\rightarrow -u ,\\
& y\rightarrow -y \; \;\;\; \mbox{and}\;\;\; \; v\rightarrow -v .
\end{split}
\ee
This operation corresponds to mirror symmetries, and the governing equations should be invariant under this operation. This invariance is valid for any crystal that has mirror symmetries along $x$ and $y$ directions. This is in particular the case for square and hexagonal (six-fold) symmetries.
For example, using the first of this symmetry transforms  Eqs.~(\ref{equv}) into
 \be
 \label{equvp}
\begin{split}
&\mathbf \rho \ddot u= -a u_x+ b u_y+cv_x - dv_y \\
&\mathbf \rho \ddot v= +a' u_x- b' u_y- c'v_x + d'v_y .
\end{split}
\ee
The equations must remain invariant, which implies that $a=d=b'=c'=0$.
Using  the second  symmetry (\ref{sym}) leads to the conclusion that  $b=c=a'=d'=0$.

In conclusion, no term with first derivatives must be present in the equations  (\ref{equv}). We have thus to consider now second derivatives.  The Taylor expansion of the functions $U$ and $V$  yields
\be
 \label{equvpp}
\begin{split}
\mathbf \rho \ddot u &= \alpha u_{xx}+ \beta u_{yy}+ \gamma v_{xy} + \tilde\alpha u_{xy}+ \tilde\beta v_{xx}+ \tilde\gamma v_{yy}\\
\mathbf \rho \ddot v &= \alpha' v_{xx}+ \beta' v_{yy}+ \gamma' u_{xy} + \tilde\alpha' v_{xy}+ \tilde\beta' u_{xx}+ \tilde\gamma' u_{yy}
\end{split}
\ee
where the coefficients $\alpha,\beta ,...$ are real constants.
As we did with first derivatives above, using the symmetry operations (\ref{sym}) leads quite simply  to the conclusion that all coefficients having tilde must be zero, getting
\be
 \label{equvppp}
\begin{split}
&\mathbf \rho \ddot u= \alpha u_{xx}+ \beta u_{yy}+ \gamma v_{xy} ,\\
&\mathbf \rho \ddot v= \alpha' v_{xx}+ \beta' v_{yy}+ \gamma' u_{xy} .
\end{split}
\ee

As stated above, the  symmetry operation (\ref{sym}) is valid for any crystal symmetry that enjoys the mirror symmetries (\ref{sym}). This includes, in two dimensions, isotropic materials, square symmetry and hexagonal crystals.

We have not yet exploited all symmetry operations. For example, for an isotropic material equations (\ref{equvppp}) must be invariant upon rotation of the coordinate system with an arbitrary angle $\theta$ around the $z$-axis. Let $x'$ and $y'$ designate the new coordinates and $u'$ and $v'$ the new displacement components after rotation, we then have
\begin{equation}
\label{rotation}
x'=x\cos\theta  - y\sin \theta,\;\;  y'=x\sin \theta  + y\cos \theta
\end{equation}
and identical relations between $(u',v')$ and $(u,v)$. The idea now is to
rewrite equations
(\ref{equvppp}) in terms of  $(x',y')$ and $(u',v')$. As a way of example we have
\begin{equation}
u_{xx}= c^2u_{x'x'}+s^2u_{y'y'}+2csu_{x'y'} ,
\end{equation}
where we have adopted the abbreviations  $\sin \theta\equiv  s$ and $\cos\theta\equiv c$. Similar relations are obtained for the other second derivatives. Then $u$ is replaced by $u=cu'+sv'$. These transformations allow one to rewrite the set of equations (\ref{equvppp}) in terms of $u'_{i'j'}$, $v'_{i'j'}$ (with $i',j'=x',y'$), and
$u'_{tt}$ and $v'_{tt}$. The transformed equations are listed in appendix \ref{set}.
The two  equations for $u'_{tt}$ and   $v'_{tt}$ have now
the form (\ref{equvpp}), not the shortest form (\ref{equvppp}).
Invariance under rotation imposes that the form of the new set of equations should be identical to (\ref{equvppp}), and this leads to relations among the coefficients. Since the invariance holds for any angle $\theta$, it holds also in particular for special values of $\theta$. For example, setting $\theta=\pi/2$ simplifies greatly the algebraic equations (see appendix \ref{set}) and this leads in a very simple way to the relations
\begin{equation}
\label{rel1}
\alpha'=\beta,\;\;  \beta'=\alpha,\;\; \gamma'=\gamma .
\end{equation}
Then, imposing that the invariance holds for any angle $\theta$ leads to
the additional relation
\begin{equation}
\gamma = \alpha-\beta ,
\label{rel2}
\end{equation}
reducing the number of independent coefficients to two.
\comment{All the other relations imposed by invariance are then automatically satisfied, meaning that the there are in total four independent relations (relations (\ref{rel1} and \ref{rel2}). }
Using these two relations allows to rewrite (\ref{equvppp}) as
\be
 \label{equvf}
\begin{split}
&\mathbf \rho \ddot u=
\beta (u_{xx}+u_{yy}) + (\alpha-\beta) (u_{xx} +v_{xy} )\\
&\mathbf \rho \ddot v=
\beta (v_{xx}+v_{yy}) + (\alpha-\beta) (v_{yy}+ u_{xy} ) ,
\end{split}
\ee
or in the more compact, vectorial form
\begin{equation}
 \label{lame2}
\mathbf \rho \ddot {\mathbf u}= \beta \nabla^2 {\mathbf u}+  (\alpha-\beta) \nabla (\nabla\cdot{\mathbf u}) ,
\end{equation}
which is nothing but the classical elastodynamics equations for an isotropic material. This concludes  our proof for an isotropic 2D material.

 Equation~(\ref{lame2}) can be rewritten as
\begin{equation}
 \label{lame3}
\rho \ddot {\mathbf u_i}= \partial _j \sigma_{ij}
\end{equation}
with
\begin{equation}
 \label{lame5}
\sigma_{ij}= 2 \beta \epsilon _{ij} + (\alpha - 2 \beta) \epsilon _{kk}\delta _{ij},
\end{equation}which is the Hooke law (recall that
$\epsilon_{ij}={1\over 2}(\partial_i {\mathbf u}_j
+ \partial_j {\mathbf u}_i)$ is the strain tensor).

In conclusion, by evoking only elementary space symmetries (and no notion of torque, or that the forces are of short range), we arrive to relation (\ref{lame3}), which implies the Cauchy principle,
i.e. that the force is localized to surface only, and Hooke law (\ref{lame5}).
In addition, the stress tensor (\ref{lame5}) is automatically symmetric. This section corresponds to the main message of our paper.
The rest is dedicated to extension to three dimensions and to other symmetries in two and three dimensions.

\section{Extension to other 2D symmetries}
Let us still consider the 2D case, but  other symmetries, namely the hexagonal and square symmetries. The same transformed equations (see appendix \ref{set}) upon a rotation by an angle $\theta$ can be used. For hexagonal symmetry, we have to impose that
Eqs. (\ref{equvppp}) are invariant for $\theta=\pi/3$.
 Simple algebraic manipulations lead to the conclusion that the six coefficients are related exactly by
relations (\ref{rel1}) and (\ref{rel2}). Therefore we get the
same conclusion as in the isotropic case. Actually it is well known \cite{Landau} in elasticity theory that in 2D the hexagonal case is identical to the isotropic one (this ceases to be valid in 3D).

For the square symmetry, invariance under a rotation $\pi/2$ (already used in the isotropic case) has led to the relations (\ref{rel1}), entailing that there are three independent coefficients. The equations for the displacement field read
\be
 \label{equvf2}
\begin{split}
&\mathbf \rho \ddot u= \alpha u_{xx}+ \beta u_{yy}+ \gamma v_{xy} ,\\
&\mathbf \rho \ddot v= \beta v_{xx}+ \alpha v_{yy}+ \gamma  u_{xy} .
\end{split}
\ee
It is a simple matter to see that these equations can be put in the form (\ref{lame3}) with
\begin{eqnarray}
\nonumber
&&\sigma_{xx}= \alpha \epsilon_{xx}+ (\gamma -\beta )\epsilon_{yy},\\
&& \sigma_{yy}= \alpha \epsilon_{yy}+ (\gamma -\beta )\epsilon_{xx}\nonumber\\
&& \sigma_{xy}=\sigma_{yx}=2\beta \epsilon_{xy}
\nonumber
\end{eqnarray}
Here again we recover the same conclusion: the  symmetry of the stress tensor is automatically
satisfied.\cite{note_stress_tensor}

\section{Extension to three dimensions}
The extension of the above approach
to three dimensions is straightforward and follows exactly the same spirit. Therefore we will keep this section brief. Let $u,v,w$ denote the three components of the displacement field. Like in 2D there are no first derivatives entering the displacement equations. There are now 18 different  second derivatives of  $u,v,w$. However several coefficients are zero. Indeed, using the 3D analogue of transformation (\ref{sym}) (which means imposing the additional
invariance with respect to the symmetry operation $z\rightarrow -z, w\rightarrow -w$) easily allows to deduce the 3D analogue of (\ref{equvppp})
\be
 \label{equvppp3D}
\begin{split}
&\mathbf \rho \ddot u= \alpha u_{xx}+ \beta u_{yy}+ \gamma v_{xy}+\delta u_{zz} + \iota w_{xz}\\
&\mathbf \rho \ddot v= \alpha' v_{xx}+ \beta' v_{yy}+ \gamma' u_{xy}+\delta' v_{zz} + \iota' w_{yz}\\
&\mathbf \rho \ddot w= \alpha'' w_{xx}+ \beta'' w_{zz}+ \gamma'' u_{xz}+\delta'' w_{yy} + \iota'' v_{yz}
\end{split}
\ee
There are a priori 15 different coefficients.
For an isotropic material we can use the invariance with respect to a rotation with an arbitrary angle $\theta$ and around an arbitrary axis (there are three independent axes). We first perform a rotation around the $z$-axis. The transformation of $(x,y)$ and $(u,v)$ are identical to (\ref{rotation}) while $z$ and $w$ are unchanged. Transforming the $(x,y,z)$ and $(u,v,w)$ into $(x',y',z')$ and $(u',v',w')$ and writing the equations for $\ddot u'$,  $\ddot v'$ and $\ddot w'$ and requiring invariance of (\ref{equvppp3D}) (see appendix \ref{set3D}) one straightforwardly  obtains the following five relations
\begin{equation}
\beta'=\alpha,\;\; \alpha'=\beta,\;\; \gamma=\gamma',\;\; \delta=\delta',\;\;\; \iota=\iota' .
\end{equation}
Using  rotations around the two other axes ($x$ and $y$) one obtains eight additional independent
relations. In total we have thus 13 relations and we are left with only two independent coefficients in equations (\ref{equvppp3D}). The reader can read off the full relations between coefficients
 just by confronting (\ref{equvppp3D}) with the following set
\be
 \label{equvppp3Dp}
\begin{split}
&\mathbf \rho \ddot u= \alpha u_{xx}+ \beta (u_{yy}+ u_{zz}) + (\alpha-\beta) (v_{xy}+w_{xz})\\
&\mathbf \rho \ddot v= \alpha v_{yy}+ \beta (v_{yy}+ v_{zz}) +  (\alpha-\beta)(u_{xy}+w_{xz})\\
&\mathbf \rho \ddot w= \alpha w_{zz}+ \beta (w_{yy}+ w_{zz}) +  (\alpha-\beta) (u_{xz}+w_{xz})
\end{split}
\ee
It is a simple matter (following exactly the same procedure as in 2D) to show that the vectorial form of this set is identical to (\ref{lame2})
and formally  the relations (\ref{lame3}) and (\ref{lame5}) still hold (with now $i,k=x,y,z$). This concludes our analysis for  a 3D isotropic material.

It is easily seen in appendix \ref{set3D} that for a cubic symmetry we are left with three independent coefficients, and the set of equations is similar to  (\ref{equvppp3Dp}) with the last coefficient
$(\alpha-\beta)$ replaced by $\gamma$. The equations can be put in the form (\ref{lame3}), with the stress tensor having the following expressions
for diagonal elements,
\begin{eqnarray}
&&\sigma_{xx}= \alpha \epsilon_{xx}+ (\gamma -\beta )(\epsilon_{yy}+\epsilon_{zz}),\nonumber\\ &&\sigma_{yy}= \alpha \epsilon_{yy}+ (\gamma -\beta )(\epsilon_{xx}+\epsilon_{zz}),\nonumber\\&&\sigma_{zz}= \alpha \epsilon_{zz}+ (\gamma -\beta )(\epsilon_{xx}+\epsilon_{yy})\nonumber
\end{eqnarray}
and for non diagonal elements
\begin{eqnarray}
&& \sigma_{ij}=\sigma_{ji}=2\beta \epsilon_{ij} .
\end{eqnarray}

That is to say, the 3D  cubic system equations can be derived fully from basic symmetry considerations.
\section{Discussion and conclusion}
The main message of this article is that for high symmetry crystals (isotropic, square and
hexagonal symmetry in 2D, isotropic and cubic symmetry in 3D) the elasticity equations can be fully derived from just evoking basic symmetry considerations. From this approach, it follows automatically that (i) the force is a divergence of a second rank tensor (stating thus that the force acts on the surface only), and (ii) that the stress tensor is symmetric. We find that this is interesting inasmuch as that these two points are usually motivated by physical principles: short range interaction for point (i) and absence of torque for point (ii). Therefore, the basic symmetries of these crystals induce naturally the symmetry of the stress tensor.

Suppose we have a cubic crystal and consider a cubic element (in the continuum spirit). In textbooks, the  force per unit surface acting on face ``$x$"
 (we mean the face having its normal along the $x$-direction) and directed along $y$, is denoted as $\sigma_{yx}$, while that acting on face ``$y$" and directed along $x$ is denoted as $\sigma_{xy}$. Since we have exploited the basic spatial symmetries, the two faces are equivalent, and therefore the equality  $\sigma_{yx}=\sigma_{xy}$, follows naturally and is encoded in the basic space symmetry. Therefore, evoking a zero torque is not only redundant, but also conveys the wrong impression that a physical principle would be needed.   Had we considered a rectangular symmetry (in the 2D case, for example),  then no equivalence between the two faces
is dictated by the spatial symmetry, and it seems necessary, at the present state of our understanding, that the zero torque condition would be required.

Note that research on elasticity theory with the aim of providing a deeper understanding  has given rise to several papers. The most relevant in the spirit of the foundation of elasticity theory are the following two research.  The work
by Dimitryev who presented  a nice  study  consisting in mapping the extended electrodynamics equations onto the theory of elasticity \cite{Dimitriev}. This work has discussed the connection between the two theories, and entirely differs from our  which raises the question of how to present the elasticity theory itself from elementary symmetries. The other noteworthy study is due to Borg \cite{Borg} who presented a combination of  dimensional analysis and symmetries  and applied it to the derivation of the strain energy. However, (i) the symmetry of the stress tensor is taken as a postulate, while here we have shown that it follows automatically from symmetries, (ii) he used the Hookes law, while here we have shown that it follows naturally from elementary space symmetries.

Finally, we would like to point out a certain redundancy in some textbooks. Indeed, as soon as one assumes that the elastic energy is a function of the deformation tensor only (actually of its invariants), then this implies automatically point (i). Indeed, let
the total elastic energy reads
\begin{equation}
F=\int f(\epsilon_{ij}) dV
\end{equation}
where  $f$ is the energy density. A variation of the energy reads
\begin{equation}
\delta F=\int {\partial f\over \partial \epsilon_{ij}}\delta \epsilon _{ij} dV=
{1\over 2}\int {\partial f\over \partial \epsilon_{ij}}
(\partial_i \delta {\mathbf u}_j+\partial_j\delta {\mathbf u}_i) dV
\end{equation}
where we have used the definition of the strain tensor $\epsilon_{ij}$ in terms of the displacement vector.
Since $i,j$ are dummy indices, one can rewrite the last term as
\begin{eqnarray}
\nonumber
&&\delta F = \int {\partial f\over \partial \epsilon_{ij} }
(\partial_i \delta {\mathbf u}_j) dV \\
&&= -\int \partial_i \left({\partial f\over \partial \epsilon_{ij}} \right)
\delta {\mathbf u}_j dV + \int \left({\partial f\over \partial \epsilon_{ij}}
\right) \delta {\mathbf u}_j n_j dA
\label{work}
\end{eqnarray}
where we have used the divergence theorem, and where $n_j$ is the normal to the elastic surface body. Actually, if we have a finite body there are also surface forces $g_i$ (like the pressure field, for example) that produce a work, given by the force times the displacement that has to be added above, $-\int g_j
\delta {\mathbf u}_j  dA$.


The first term on the right hand side of  (\ref{work}) is a product of some entity, $\partial_i ({\partial f\over \partial \epsilon_{ij}})$, times a displacement, but this is nothing but the very definition of the work done by internal stresses. In other words $\partial_i ({\partial f\over \partial \epsilon_{ij}})$ is the force, which naturally reads as the divergence of a second order tensor  $\sigma_{ij}\equiv {\partial f\over \partial \epsilon_{ij}}$. At equilibrium  the total variation vanishes. Thus, the bulk term in (\ref{work}) vanishes (corresponding to $\nabla\mbox{\boldmath{$\sigma$}} =0$) as well as  the surface term, leading to $g_i=\sigma_{ij}n_j$, which is nothing but the boundary condition at the external surface.

\noindent {\bf Acknowledgement}: C.M. Acknowledges financial support from CNES (Centre National d'Etudes Spatiales) and ESA (European Space Agency).
\begin{appendix}
\section{The 2D case}
\label{set}
Using the transformation (\ref{rotation}) the set of equations (\ref{equvppp}) transforms into
\be
\label{setu'}
\begin{split}
\rho(c\ddot u'+s\ddot v')=& A u'_{x'x'}+ B u'_{y'y'}+ C v'_{x'x'}\\
& + D v'_{y'y'}+E u'_{x'y'}+ F v'_{x'y'}\\
\rho(c\ddot v'-s\ddot u')=& A' u'_{x'x'}+ B' u'_{y'y'}+ C' v'_{x'x'}\\
& + D' v'_{y'y'}+E' u'_{x'y'}+ F' v'_{x'y'}
\end{split}
\ee
where
\be
\begin{split}
&A = (\alpha c^2+\beta s^2)c+\gamma cs^2,\;\; \\
&B=(\alpha s^2+\beta c^2)c-\gamma s^2c\nonumber \\
&C = (\alpha c^2+\beta s^2)s-\gamma c^2s,\;\; \\
& D=(\alpha s^2+\beta c^2)s+\gamma c^2s \nonumber\\
&E = 2(\alpha-\beta)c^2s-s(c^2-s^2)\gamma, \;\;\ \\
& F=2(\alpha-\beta)cs^2 + c(c^2-s^2)\gamma \nonumber\\
& A' = -(\alpha' c^2+\beta' s^2)s-\gamma' c^2s,\;\; \\
& B'=-(\alpha' s^2+\beta' c^2)s + \gamma' c^2s,\;\\
& C'=(\alpha' c^2+\beta' s^2)c-\gamma' cs^2,\; \\
& D'=(\alpha' s^2+\beta' c^2)c+\gamma' cs^2, \;\;\\
& E'=2(\beta'-\alpha')cs^2+c(c^2-s^2)\gamma' ,\;\;\nonumber \\
&F'=2(\alpha'-\beta')c^2s+s(c^2-s^2)\gamma'
\end{split}
\ee
Multiplying the first (resp. second) equation in (\ref{setu'}) by
$c\equiv \cos\theta$ (resp. $s\equiv\sin\theta$) and the second by $s$ (resp. $c$) and making the difference (resp. the sum), one obtains the equations for $\ddot u'$ and $\ddot v'$:
\be
\label{setu'p}
\begin{split}
\rho\ddot u' &= (cA -sA') u'_{x'x'}+ (cB-sB') u'_{y'y'}\\
& +( cC-sC') v'_{x'x'}+(c D-sD') v'_{y'y'}\\
& +(cE-sE') u'_{x'y'}+ (cF-sF') v'_{x'y'}\\
\rho\ddot v' &= (sA +cA') u'_{x'x'}+ (sB+cB') u'_{y'y'}\\
& +( sC+cC') v'_{x'x'}+(s D+cD') v'_{y'y'}\\
& +(sE+cE') u'_{x'y'}+ (sF+cF') v'_{x'y'}
\end{split}
\ee
Comparison with (\ref{equvppp}) tells us that the invariance is preserved if and only and if
we have the following relations
\be
\label{setu'pp}
\begin{split}
& cA -sA'=\alpha,\;\; cB-sB'=\beta,\;\; cF-sF'=\gamma,  \\
& cC-sC'=0,\;\; c D-sD',\;\; cE-sE'=0\\
& sC+cC'=\alpha',\;\; s D+cD'=\beta',\;\;sE+cE'=\gamma',\\
&  sA +cA'=0,\;\; sB+cB'=0,\;\; sF+cF'=0
\end{split}
\ee
For a square (as well as an isotropic) crystal this relations holds for $\theta=\pi/2$. The algebra is quite simple (since $c=0$, $s=1$), and we immediately obtain (\ref{rel1}). This means that for a square symmetry only relation (\ref{rel1}) holds. For an isotropic material we still can impose that the set (\ref{setu'pp}) holds for any $\theta$ (by using the already obtained result (\ref{rel1}), which greatly simplifies the algebra). This yields relation (\ref{rel2}). Finally for an hexagonal symmetry,
we must impose invariance with respect to $\theta=\pi/3$ ($c=1/2$ and $s=\sqrt{3}/2)$.
Reporting into (\ref{setu'pp}) leads to quite simple algebraic equations leading to relations (\ref{rel1}) and (\ref{rel2}). In other words, the hexagonal symmetry obeys (in two dimensions) exactly the same set of equations as isotropic materials, a well known fact \cite{Landau}.

\section{The 3D case}
\label{set3D}
The 3D case presents no special challenge, and the reasoning is exactly the same as in 2D.  We start from (\ref{equvppp3D}) and performs a rotation around the $z$-axis. In this case $(x',y',z')=(cx-sy,sx+cy,z)$ and identical relations between the displacement components. The transformed equation in the new coordinate system reads
\be
\label{setu'3D}
\begin{split}
\rho(c\ddot u'+s\ddot v') &= T_{2D} + \delta (cu'_{z'z'}+sv'_{z'z'})\\
& + \iota (cw'_{x'z'}+sw'_{y'z'})\\
\rho(c\ddot v'-s\ddot u') &= T'_{2D}+ \delta' (cv'_{z'z'}-su'_{z'z'})\\
& + \iota' (cw'_{y'z'}-sw'_{x'z'})\\
 \rho\ddot w' &= \alpha''(c^2w'_{x'x'}+s^2 w'_{y'y'}+2csw'_{x'y'})\\
&+\beta'' w'_{z'z'}+ \gamma'' (c^2u'_{x'z'}+csv'_{x'z'}+csu'_{y'z'}+s^2v'_{y'z'})\\
&+\iota''(c^2v'_{y'z'}-csv'_{x'z'}-csu'_{y'z'}+s^2u'_{x'z'})\\
&+ \delta''(s^2 w'_{x'x'}+c^2w'_{y'y'}-2csw'_{x'y'})
\end{split}
\ee
where $T_{2D}$ and $T'_{2D}$ are the right hand sides  of the first and second equations in (\ref{setu'}), respectively. The next step is to perform the same manipulations done for the 2D case (i.e. extract equations for  $\ddot u'$ and $\ddot v'$;  that for  $\ddot w'$ is already explicit in the set (\ref{setu'3D})).
Imposing invariance of obtained equations (i.e. that they must have the form (\ref{equvppp3D})) yields the same relations (\ref{rel1}) and (\ref{rel2}) (obtained in 2D) plus two new independent relations (implied by the third variable $w$):
\begin{equation}
\label{rel3}
\delta=\delta',\;\; \iota=\iota' .
\end{equation}
Performing rotations about the other two axes $x$ and $y$ axes yield new additional relations among coefficients. Actually, it is even much simpler to deduce the new relations, just by successively exchanging the role of $w$ and $u$ and then the role of $w$ and $v$ (without making any algebraic calculation). For example the above equality $\delta=\delta'$ means (when inspecting the set (\ref{equvppp3D})) that the coefficient $u_{zz}$ in the $u$-equation is equal to the coefficient of $v_{zz}$ in the $v$-equation. This obviously implies that the coefficient of $w_{xx}$ in the $w$-equation should be equal to the coefficient of $v_{xx}$ of the $v$-equation. For the same reason, the coefficient of $w_{yy}$ in the $w$-equation should be equal to the coefficient of $u_{yy}$ of the $u$-equation. Similar reasoning can be made for the relation $\iota=\iota'$. These considerations lead to
\begin{equation}
\label{rel4}
 \delta''=\delta=\beta,\;\; \iota''=\iota=\gamma,\;\; \gamma''=\gamma .
 \end{equation}
 Using the relations (\ref{rel1}), (\ref{rel2}), (\ref{rel3}) and (\ref{rel4}), leads to the final system  (\ref{equvppp3Dp}). For the cubic case, the relation (\ref{rel2}) (which follows by imposing an arbitrary rotation angle) does not hold and the equations are obtained by replacing in (\ref{equvppp3Dp})
the last coefficients $(\alpha-\beta)$ by $\gamma$.
 \end{appendix}


\begin{thebibliography}{99}
\bibitem{note_notation}
For ease of notations, $u_i$ will indicate a spatial derivative, see the next Section.
\bibitem{note_Lame}
The more common Young modulus $E$ and Poisson ratio $\nu$ are algebraic
combinations of $\lambda$ and $\mu$:
$E=\mu (3\lambda + 2\mu)/(\lambda + \mu )$ and
$\nu =\lambda /2(\lambda + \mu )$.
\bibitem{note_long_range}
We are assuming here that forces are local.
\bibitem{note_stress_tensor}
More generally, we can show that Eqs.~(\ref{equvppp})  can always be written in the form
(\ref{lame3}). However, the resulting stress tensor is symmetric if and only if $\alpha'=\beta$.
\end{thebibliography}
  \end{document}